\documentclass{ieeeaccess}

\usepackage{cite}
\usepackage{amsmath,amssymb,amsfonts}
\usepackage{algorithmic}
\usepackage{algorithm}
\usepackage{graphicx}
\usepackage{amsmath}  
\usepackage{makecell} 
\usepackage{graphicx} 
\usepackage{multirow} 
\usepackage{textcomp}
\usepackage{pifont}
\usepackage{algorithm}
\usepackage{float}
\usepackage{cite}        
\usepackage{graphicx}    
\usepackage{amsmath}     
\usepackage{amssymb}
\usepackage{url}         
\def\BibTeX{{\rm B\kern-.05em{\sc i\kern-.025em b}\kern-.08em
    T\kern-.1667em\lower.7ex\hbox{E}\kern-.125emX}}
\begin{document}
\history{Date of publication xxxx 00, 0000, date of current version xxxx 00, 0000.}
\doi{10.1109/ACCESS.2017.DOI}

\title{Adaptive Network Selection for Latency- Aware V2X Systems under Varying Network and Vehicle Densities}
\author{\uppercase{Muhammad Z. Haq}\authorrefmark{1}, 
\uppercase{Nadia N. Qadri}\authorrefmark{2}, \IEEEmembership{Senior Member, IEEE},
\uppercase{Omer Chughtai}\authorrefmark{2}, \IEEEmembership{Senior Member, IEEE},
\uppercase{Sadiq A. Ahmad}\authorrefmark{1}, \IEEEmembership{Member, IEEE},
\uppercase{Waqas Khalid\authorrefmark{3},  \IEEEmembership{Member, IEEE}, and Heejung Yu}\authorrefmark{4},
\IEEEmembership{Senior Member, IEEE}}
\address[1]{Department of Electrical Engineering, COMSATS University Islamabad (CUI), Wah Campus, Wah Cantt, Pakistan; ziaulhaq\_cui@ciitwah.edu.pk, engrsadiqahmad@ciitwah.edu.pk}
\address[2]{Department of Computer Engineering, CUI Wah Campus, Wah Cantt, Pakistan; drnadia@ciitwah.edu.pk, omer.chughtai@cuiwah.edu.pk}
\address[3]{Institute of Industrial Technology, Korea University, Sejong, 30019, South Korea; waqas283@gmail.com}
\address[4]{Department of Electronics and Information Engineering, Korea University, Sejong, 30019, South Korea; heejungyu@korea.ac.kr}

\newdimen\xfigwd

\tfootnote{This work was supported by Korea University Grant.}

\markboth
{Author \headeretal: Preparation of Papers for IEEE Access Journal}
{Author \headeretal: Preparation of Papers for IEEE Access Journal}

\corresp{Corresponding authors: Heejung Yu and Waqas Khalid}

\begin{abstract}

\textcolor{red}{This paper presents ANS-V2X, an Adaptive Network Selection framework tailored for latency-aware V2X systems operating under varying vehicle densities and heterogeneous network conditions. Modern vehicular environments demand low-latency and high-throughput communication, yet real-time network selection is hindered by diverse application requirements and the coexistence of multiple Radio Access Technologies (RATs) such as 4G, 5G, and ad hoc links. ANS-V2X employs a heuristic-driven approach to assign vehicles to networks by considering application sensitivity, latency, computational load, and directionality constraints. The framework is benchmarked against a Mixed-Integer Linear Programming (MILP) formulation for optimal solutions and a Q-learning-based method representing reinforcement learning. Simulation results demonstrate that ANS-V2X achieves near-optimal performance, typically within 5 to 10\% of the utility achieved by MILP-V2X, while reducing execution time by more than 85\%. Although MILP-V2X offers globally optimal results, its computation time often exceeds 100 milliseconds, making it unsuitable for real-time applications. The Q-learning-based method is more adaptable but requires extensive training and converges slowly in dynamic scenarios. In contrast, ANS-V2X completes decisions in under 15 milliseconds and consistently delivers lower latency than both alternatives. This confirms its suitability for real-time, edge-level deployment in latency-critical V2X systems.}

\end{abstract}


\titlepgskip=-15pt

\maketitle

\section{Introduction}
\label{sec:intro}
The evolution of Intelligent Transportation Systems (ITS) has made Vehicle-to-Everything (V2X) communication essential for improving road safety, supporting autonomous mobility, and enabling infotainment services. V2X includes several communication paradigms such as Vehicle-to-Vehicle (V2V), Vehicle-to-Infrastructure (V2I), Vehicle-to-Pedestrian (V2P), and Vehicle-to-Network (V2N). These modes of communication collectively contribute to cooperative driving, reduction of traffic congestion, and delivery of context-aware mobility and infotainment applications~\cite{pawar2024intelligent}. \textcolor{red}{The effectiveness of V2X communication relies on low-latency, reliable data exchange, which is crucial for making real-time decisions in dynamic traffic environments. Recent advancements in edge computing and artificial intelligence are further enhancing the adaptability and responsiveness of V2X systems, as also discussed in \cite{rahmani2025optimizing}.} Advancements in wireless technologies such as 5G and the emerging 6G are significantly enhancing V2X capabilities. However, integrating multiple Radio Access Technologies (RATs), including Dedicated Short-Range Communications (DSRC), Cellular V2X (C-V2X), Wi-Fi, and 5G New Radio (NR)-V2X, introduces challenges related to latency, mobility, and service reliability. Additionally, the evolution of V2X has progressed beyond traditional wireless technologies to incorporate more advanced cellular networks. 5G technology, with its support for ultra-reliable low-latency communication (URLLC) and enhanced mobile broadband (eMBB), is well-suited for real-time V2X applications such as cooperative collision avoidance and vehicle platooning~\cite{adhikari2024comparative}. In the future, 6G networks are expected to further enhance these capabilities with features such as terahertz communication and holographic telepresence~\cite{othman2025key}. 

Despite these advancements, the integration of diverse communication technologies presents significant challenges. V2X systems now rely on a heterogeneous mix of RATs, including DSRC, C-V2X, Wi-Fi (IEEE 802.11p/ac/ax), and 5G NR-V2X, each offering different performance characteristics in terms of coverage, latency, and reliability. This diversity complicates efforts to ensure seamless and consistent connectivity, particularly in environments with fluctuating network conditions~\cite{10749940}. Moreover, the increased connectivity of vehicles has introduced cybersecurity concerns. As V2X enables critical applications like emergency vehicle warnings and intersection movement assistance, safeguarding data integrity and user privacy is paramount ~\cite{muslam2024enhancing}. Despite the growing importance of vehicular connectivity in Vehicular Ad-hoc Networks (VANETs), maintaining stable and reliable communication remains a significant challenge due to the inherently high mobility and dynamic topology of vehicles. These conditions often lead to intermittent connections, increased latency, and reduced communication reliability, which in turn hinder the effectiveness of V2X communications~\cite{souri2024systematic}.

To address these challenges, an adaptive and efficient network selection mechanism is essential for ensuring optimal connectivity in dynamic vehicular environments. This paper focuses on enhancing V2X communication systems by integrating multiple RATs and developing a framework that dynamically allocates network resources based on the specific needs of V2X applications.

\subsection{Problem Statement}

The diverse and evolving nature of V2X applications imposes a wide range of Quality of Service (QoS) requirements. Safety-critical applications, such as forward collision warnings or emergency braking alerts, demand ultra-low latency and high reliability. Conversely, infotainment applications prioritize high bandwidth and sustained throughput. Traditional static or rule-based network selection fails to meet the dynamic QoS demands in V2X environments. This leads to suboptimal application performance, especially in high-mobility and variable-density scenarios.

An adaptive, real-time, application-aware network selection mechanism is essential to ensure seamless V2X communication. This paper proposes ANS-V2X to bridge this gap by selecting networks based on latency sensitivity, application type, network condition, and mobility context.

\subsection{Motivation}

The motivation for this work stems from the growing necessity of adaptive connectivity in the V2X domain. To support autonomous mobility and improve safety, an efficient and reliable network selection framework must be capable of managing not only signal strength and coverage but also factors such as vehicular mobility, application urgency, and network congestion. The dynamic nature of vehicular environments, particularly with vehicles transitioning between various network coverage zones, necessitates a system that can respond in real-time to ensure continuous, reliable communication. By developing a network selection framework that prioritizes low-latency connectivity for safety-critical services and high-bandwidth access for infotainment, this work aims to ensure optimal connectivity and improved V2X performance. This adaptive approach will be essential for the scalability of autonomous mobility systems and the successful deployment of V2X infrastructure in real-world applications.



\subsection{Contributions}

This work addresses the challenges associated with dynamic and application-aware network selection in V2X communication systems. A major contribution is the development of a unified framework that integrates heterogeneous RATs, including DSRC, Long-Term Evolution (LTE), and 5G NR. \textcolor{red}{Such integration enhances spectral efficiency, improves link reliability, and extends connectivity in diverse vehicular scenarios.} In addition, the proposed framework adopts an application-aware and adaptive network selection strategy, where the most suitable network is chosen based on real-time assessment of application requirements, such as those needed for safety versus infotainment, along with current network conditions. This ensures that QoS is maintained in a context-sensitive manner. \textcolor{red}{Unlike many previous studies that assume vertical handover without explicitly addressing its complexities, this work incorporates a mechanism that enables seamless inter-RAT handovers, accounting for latency overhead and minimizing packet loss during mobility.} \textcolor{red}{Moreover, the proposed ANS-V2X algorithm addresses the heterogeneous demands of V2X applications by dynamically aligning service priorities with available network capabilities, delivering robust and efficient connectivity under varying network and vehicular conditions.}

The rest of the paper is organized: Section II presents relevant literature, while Section III outlines the detailed problem definition. Section IV introduces the system model. Section V encompasses the problem formulation and discusses the simulation results. Finally, Section VI concludes the paper.

\section{Literature Review}
\label{sec:litreview}

Recent advancements in V2X communications have attracted considerable attention as vehicles progress towards greater autonomy and interconnectivity. One of the key challenges in this domain is ensuring seamless, reliable, and efficient communication in dynamic vehicular environments. This challenge has led to numerous studies exploring various aspects of V2X, including radio access technology integration, network selection mechanisms, and QoS optimization.

V2X communications leverage a variety of RATs, each offering distinct capabilities and trade-offs. While IEEE 802.11p-based DSRC has been used for its low-latency, peer-to-peer nature, its limitations in scalability and reliability under dense traffic conditions have driven a shift toward C-V2X, which employs LTE and 5G NR for enhanced coverage, quality of service, and centralized coordination~\cite{kanavos2025v2x}. With the advent of 5G NR, particularly from 3GPP Release 16 onward, features such as URLLC, sidelink communications, and network slicing have enabled differentiated service levels for safety-critical and infotainment V2X applications~\cite{clancy2024wireless}. These enhancements have positioned C-V2X as a core enabler of advanced vehicular use cases. Looking ahead, 6G is expected to deliver transformative capabilities, such as integrated sensing and communication (ISAC), joint communication-computation paradigms, and holographic-type communications, which will further increase network complexity and raise the bar for dynamic resource and RAT management~\cite{lu2024generative}.

Traditional network selection approaches in vehicular environments have largely relied on static policies or simple metrics such as Received Signal Strength Indicator (RSSI), which fall short in dynamic scenarios where network conditions and application demands fluctuate rapidly~\cite{li2013intelligent}. Recent research has explored context-aware and adaptive network selection mechanisms that consider a broader set of parameters, including application type, vehicular speed, latency requirements, and network load. Multi-Criteria Decision-Making (MCDM) techniques, such as the Analytic Hierarchy Process (AHP), TOPSIS, and variants of fuzzy logic, have been utilized to prioritize network selection metrics dynamically~\cite{gerla2014internet, gupta2022fuzzy}. However, these techniques are often computationally intensive and may lack the responsiveness required for real-time vehicular mobility.

Vertical handover, defined as the transition between heterogeneous RATs (e.g., from DSRC to C-V2X), is a key mechanism for ensuring continuous and reliable connectivity in V2X. Several mobility-aware handover strategies have been proposed that use GPS trajectory prediction, link quality estimation, and predictive analytics to optimize handover timing and reduce service disruption~\cite{jiang2023adaptive}. Despite these advances, challenges remain in achieving seamless, low-latency handovers at high vehicular speeds without compromising packet delivery or application QoS.

Security and privacy remain critical considerations in network selection, especially as V2X expands into safety-critical applications. Cryptographic frameworks, identity-based authentication, and trust models have been developed to protect vehicular communication from spoofing, replay attacks, and unauthorized access~\cite{petit2015potential}. Yet, these security mechanisms often operate independently of the underlying network selection logic, leading to a lack of coordination between connectivity management and security provisioning.

The trend toward heterogeneous network integration is accelerating, with recent studies investigating cooperative perception, edge-assisted offloading, and AI-enhanced communication strategies. Emerging techniques such as semantic-aware data transmission, ISAC, and generative models for adaptive compression are being explored to support bandwidth-efficient and resilient V2X services~\cite{krayani2023integrated}. Nonetheless, most existing approaches focus on improving communication or perception independently, while the challenge of real-time, application-aware, and seamless network selection across diverse RATs remains inadequately addressed. While 3GPP Release 16 standards have introduced multi-connectivity and resource allocation mechanisms, they lack the real-time responsiveness and context-aware optimization required for diverse V2X applications. Therefore, despite substantial advancements in V2X connectivity, the literature still lacks a cohesive solution that combines multi-RAT integration, real-time adaptability, application sensitivity, and seamless vertical handovers.

Table \ref{tab:litreview} summarizes key studies in the field of V2X communications, with a focus on application-aware adaptive network selection, vertical handovers, and real-time adaptability. \textcolor{red}{Additionally, it summarizes key limitations of recent works, highlighting the need for an adaptive and real-time V2X connectivity framework}. The studies are evaluated based on several critical factors, including the type of communication technologies used (e.g., DSRC, C-V2X, 5G NR), the support for safety and infotainment applications, and the presence of mechanisms for vertical handovers and real-time adaptability. This comparative analysis helps to highlight the existing gaps in the literature and motivates the proposed work, which aims to address these challenges by introducing an adaptive network selection framework for seamless V2X communication. \textcolor{red}{Recent studies have explored AI-based and optimization-driven solutions for network selection in V2X and edge computing systems. For instance, \cite{liu2021robust} proposes a robust decision framework under uncertain vehicular dynamics, while \cite{yang2022multiaccess} uses deep reinforcement learning for access selection in multi-RAT vehicular environments. Other works, such as \cite{zhang2021mobilityaware} incorporate mobility prediction into network handoff decisions to improve stability and performance. Compared to these approaches, ANS-V2X emphasizes low-latency, real-time adaptability without requiring pretraining or heavy computation.}


\begin{table*}[hbt]
\centering
\caption{Updated Summary of Literature and Research Gaps in Application-Aware V2X Connectivity}
\label{tab:litreview}
\renewcommand{\arraystretch}{1.2}
\begin{tabular}{|p{4.0cm}|c|c|c|c|c|c|c|c|}
\hline
\multicolumn{1}{|c|}{\textbf{Research Emphasis (Ref.)}} & 
\textbf{DSRC} & 
\makecell{\textbf{C-V2X} \\ \textbf{/5G}} & 
\makecell{\textbf{Safety} \\ \textbf{App.}} & 
\makecell{\textbf{Infotain.} \\ \textbf{App.}} & 
\makecell{\textbf{App-Aware} \\ \textbf{Sel.}} & 
\makecell{\textbf{Vertical} \\ \textbf{Handover}} & 
\makecell{\textbf{Real-time} \\ \textbf{Adapt.}} & 
\makecell{\textbf{Explicit} \\ \textbf{HO Mgmt.}} \\ \hline

DSRC Communication~\cite{kenney2011dedicated} & \checkmark & \ding{55} & \checkmark & \ding{55} & \ding{55} & \checkmark & \ding{55} & \ding{55} \\ \hline

C-V2X Interworking~\cite{kanavos2025v2x} & \checkmark & \checkmark & \checkmark & \checkmark & \ding{55} & \checkmark & \ding{55} & \ding{55} \\ \hline

5G NR Advancement~\cite{clancy2024wireless} & \ding{55} & \checkmark & \checkmark & \checkmark & \ding{55} & \checkmark & \ding{55} & \ding{55} \\ \hline

Mobility-Aware Handover~\cite{jiang2023adaptive} & \ding{55} & \checkmark & \checkmark & \ding{55} & \ding{55} & \checkmark & \checkmark & \checkmark \\ \hline

MCDM Methods~\cite{gerla2014internet} & \ding{55} & \checkmark & \checkmark & \checkmark & \checkmark & \checkmark & \checkmark & \ding{55} \\ \hline

Generative Semantic Communication~\cite{lu2024generative} & \ding{55} & \checkmark & \ding{55} & \checkmark & \ding{55} & \checkmark & \ding{55} & \ding{55} \\ \hline

Security in V2X~\cite{petit2015potential} & \checkmark & \checkmark & \checkmark & \ding{55} & \ding{55} & \checkmark & \ding{55} & \ding{55} \\ \hline

GPS Spoof/Jam Detection~\cite{krayani2023integrated} & \ding{55} & \checkmark & \checkmark & \ding{55} & \ding{55} & \checkmark & \ding{55} & \ding{55} \\ \hline

Danger-Aware Networking~\cite{sadeeq2024research} & \ding{55} & \checkmark & \checkmark & \ding{55} & \checkmark & \checkmark & \ding{55} & \ding{55} \\ \hline

\textbf{Proposed Work} & 
\textcolor{red}{\checkmark} & 
\textcolor{red}{\checkmark} & 
\textcolor{red}{\checkmark} & 
\textcolor{red}{\checkmark} & 
\textcolor{red}{\checkmark} & 
\textcolor{red}{\checkmark} & 
\textcolor{red}{\checkmark} & 
\textcolor{red}{\checkmark} \\ \hline

\end{tabular}
\end{table*}










\subsection{Research Gaps}

Despite advancements in V2X communications, a unified framework that integrates heterogeneous RATs, supports real-time, application-aware network selection, and ensures seamless vertical handovers is still lacking. \textcolor{red}{Most existing approaches are \textcolor{red}{narrow in scope}, focusing on individual protocol layers (e.g., physical or network) and exhibiting limited scalability in dynamic, resource-constrained vehicular environments.}

Key research gaps include:

\begin{itemize}
    \item Integration of Heterogeneous RATs: Few solutions \textcolor{red}{effectively} integrate technologies such as DSRC, C-V2X, and 5G to adapt to diverse network conditions and application requirements.

    \item Real-time Application-Aware Network Selection: Existing methods typically lack the ability to \textcolor{red}{dynamically respond to real-time application demands}, especially when prioritizing safety-critical services over infotainment.

    \item Seamless Vertical Handover: While vertical handovers across multiple RATs have been explored, challenges remain in ensuring \textcolor{red}{fast and lossless transitions}, particularly when balancing latency-sensitive and throughput-intensive services.

    \item Scalability and Adaptability: Existing frameworks \textcolor{red}{often lack robustness in handling fast-changing topologies and network variability} in highly mobile, resource-limited V2X environments.
\end{itemize}

The proposed work aims to address these gaps by offering a unified framework that dynamically evaluates vehicular application requirements and network conditions. This framework performs intelligent resource allocation across DSRC, LTE, 5G NR, and Wi-Fi networks, ensuring robust, low-latency connectivity for safety-critical services while providing high-throughput access for infotainment applications. By incorporating adaptive decision-making and seamless handover mechanisms, the proposed solution enhances the continuity and reliability of V2X communication, which is critical for next-generation autonomous and connected mobility systems. \textcolor{red}{To bridge these gaps, there is a need for a lightweight, adaptive network selection framework that ensures context-aware, low-latency connectivity while supporting seamless vertical handovers across heterogeneous access technologies.} 

\section{System Model and Problem Formulation}

\begin{figure*}[!htb]
  \centering
  \includegraphics[width=\textwidth]{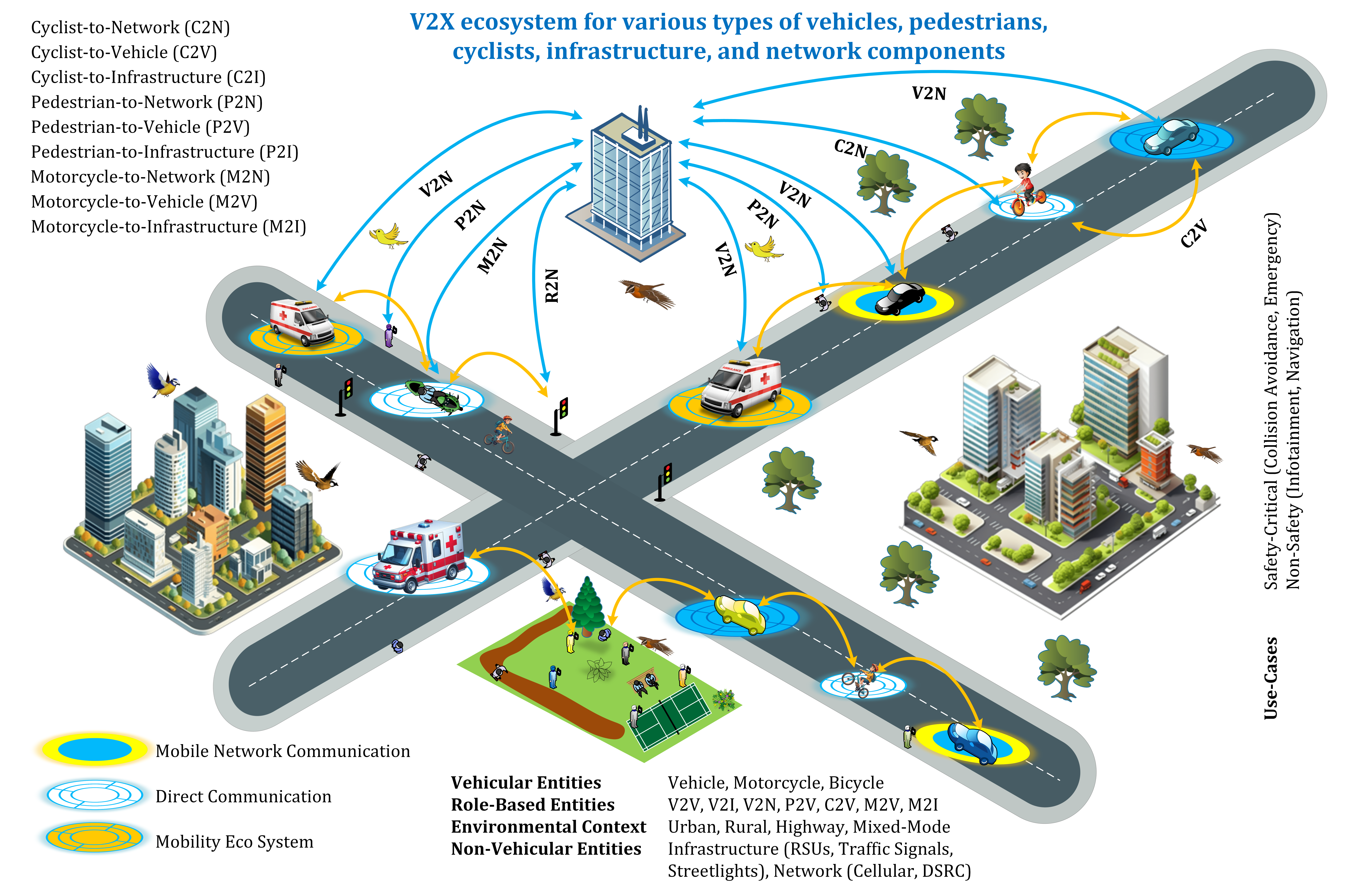}
  \caption{System model of the V2X communication paradigm showing V2V, V2I, V2P, and V2N interactions for seamless connectivity across heterogeneous networks.} 
  \label{Fig:v2X_SysModel}
\end{figure*}

The framework illustrated in Figure \ref{Fig:v2X_SysModel} represents the heterogeneous V2X communication environment targeted in this work. It comprises multiple communication paradigms, including Vehicle-to-Vehicle (V2V), Vehicle-to-Infrastructure (V2I), Vehicle-to-Pedestrian (V2P), and Vehicle-to-Network (V2N) links. Each mode supports specific types of data exchanges critical for vehicular safety, traffic management, and infotainment. The proposed resource allocation model, formulated as an integer optimization problem and solved using a branch-and-bound-based method, is integrated within this broader V2X landscape. Although the figure shows all V2X components for completeness, the optimization mechanism focuses specifically on the dynamic selection and allocation of communication resources for safety and infotainment services delivered over multiple available wireless networks. This holistic view highlights the interaction among different V2X entities, emphasizing the system-level significance of optimal service distribution across heterogeneous links.

In vehicular networks, vehicles often operate within the overlapping coverage areas of multiple heterogeneous wireless networks, such as LTE (4G), 5G, and ad hoc networks (e.g., DSRC or Wi-Fi-based VANETs). These networks differ in terms of coverage area, bandwidth, latency, signal strength, and computational resources. Selecting the optimal network for V2X communication is crucial, particularly for supporting diverse applications with varying QoS requirements. For instance, safety-critical applications like collision avoidance demand low latency and high reliability, while infotainment services (e.g., video streaming) require high throughput.

The primary research challenge lies in dynamically assigning the most suitable network to each vehicle in real-time, given the fluctuating availability of resources and mobility of vehicles. This research aims to formulate and solve a joint optimization problem that selects the optimal network for each vehicle, minimizing the total data dissemination time across all vehicles, while satisfying resource, directionality, power, and QoS constraints.

Let $\mathcal{V} = {1, 2, \ldots, V}$ be the set of vehicles, and $\mathcal{N} = {1, 2, \ldots, N}$ be the set of candidate networks, where $i=0$ represents the currently connected network prior to any reassignment. Define the binary decision variable $\alpha_{ij} \in {0,1}$ to indicate whether vehicle $\nu_j$ is assigned to network $n_i$.

\subsection{Objective Function}

The objective is to minimize the total data dissemination time for all vehicles:

\begin{equation} \label{eq:main_obj}
\underset{\boldsymbol{\alpha}, \mathbf{b}, \mathbf{f}}{\text{min}} \sum_{j=1}^{V} T_j
\end{equation}
where $T_j$ is the dissemination time for vehicle $\nu_j$, defined as:

\begin{equation} \label{eq:Tj}
T_j = \sum_{i=0}^{N} \alpha_{ij} T_{ij}
\end{equation}

Here, $T_{ij}$ is the total delay experienced by vehicle $\nu_j$ when connected to network $n_i$, given by:

\begin{equation} \label{eq:Tij}
T_{ij} = T_{ij}^{\text{trans}} + T_{ij}^{\text{comp}} + T_{ij}^{\text{handover}}
\end{equation}

The transmission time $T_{ij}^{\text{trans}}$, computation time $T_{ij}^{\text{comp}}$, and handover delay $T_{ij}^{\text{handover}}$ are computed as:

\begin{equation} \label{eq:Ttrans}
T_{ij}^{\text{trans}} = \frac{s_j}{r_{ij}} + \Delta_i + \frac{P_{ij}^{\text{idle}} \cdot \sigma + P_{ij}^c \cdot T_{ij}^c}{P_{ij}^s}
\end{equation}

\begin{equation} \label{eq:Tcomp}
T_{ij}^{\text{comp}} = \frac{c_j}{f_{ij}}
\end{equation}

\textcolor{red}{A switching delay term $T_{ij}^{\text{handover}}$ is included if vehicle $\nu_j$ switches from its current network to $n_i$. This models the inter-RAT switching latency and is defined by a Gaussian distribution with a mean of 20 ms and a 5 ms standard deviation \cite{mahmood2019handover}. If no network change occurs (i.e., when $i = 0$\textbf), then $T_{ij}^{\text{handover}} = 0$.}

Substituting \eqref{eq:Ttrans} and \eqref{eq:Tcomp} into \eqref{eq:Tij}, we get:

\begin{equation} \label{eq:Tij_expanded}
T_{ij} = \frac{s_j}{r_{ij}} + \Delta_i + \frac{P_{ij}^{\text{idle}} \cdot \sigma + P_{ij}^c \cdot T_{ij}^c}{P_{ij}^s} + \frac{c_j}{f_{ij}} + T_{ij}^{\text{handover}}
\end{equation}

Thus, the objective function becomes:

\begin{align} \label{eq:final_obj}
\min_{\alpha_{ij}, b_{ij}, f_{ij}} \sum_{j=1}^{V} \sum_{i=0}^{N} \alpha_{ij} \Bigg( 
    & \underbrace{\frac{s_j}{r_{ij}} + \Delta_i}_{\text{Transmission time}} 
    + \underbrace{\frac{P_{ij}^{\text{idle}} \cdot \sigma + P_{ij}^c \cdot T_{ij}^c}{P_{ij}^s}}_{\text{Power-Induced time}} \nonumber \\
    & + \underbrace{\frac{c_j}{f_{ij}}}_{\text{Computation time}} 
    + \underbrace{T_{ij}^{\text{handover}}}_{\text{Handover delay}} 
\Bigg)
\end{align}

\noindent
Where:
\begin{itemize}
  \item $s_j$: Data size of vehicle $\nu_j$ (in bits).
  \item $r_{ij}$: Data rate between vehicle $\nu_j$ and network $n_i$.
  \item $\Delta_i$: Network latency.
  \item $P_{ij}^{\text{idle}}$: Idle power consumption of network $n_i$ with vehicle $\nu_j$.
  \item $P_{ij}^c$: Computational power of network $n_i$.
  \item $T_{ij}^c$: Computation time at network $n_i$.
  \item $P_{ij}^s$: Stable transmission power.
  \item $c_j$: Computational workload required by vehicle $\nu_j$.
  \item $f_{ij}$: Computation capacity of network $n_i$ assigned to vehicle $\nu_j$.
  \item $T_{ij}^{\text{handover}}$: Handover latency when switching from the current network to $n_i$.
\end{itemize}

\subsection{Constraints}

\subsubsection{Bandwidth Constraint}
\begin{equation} \label{eq:BW}
\sum_{j=1}^{V} \alpha_{ij} b_{ij} \leq B_i, \quad \forall i \in \mathcal{N}
\end{equation}
The bandwidth constraint ensures that the sum of the bandwidth requirements for each vehicle to connect to a specific network does not exceed the total available bandwidth of that network. $B_i$ denotes the total available bandwidth in network $n_i$, representing the maximum bandwidth the network can provide for all connected vehicles. $b_{ij}$ is the bandwidth required for vehicle $\nu_j$ to connect to network $n_i$.

\subsubsection{Network Selection Constraint}

Each vehicle must be connected to one network at a time, and the total sum of binary variables $\alpha_{ij}$ over all networks should not exceed one. This constraint ensures that each vehicle connects to at most one network:

\begin{equation} \label{eq:net_sel}
\sum_{i=0}^{N} \alpha_{ij} = 1, \quad \forall j \in \mathcal{V}
\end{equation}
where $\mathcal{V}$ is the set of vehicles and $N$ represents the total number of available networks.

\subsubsection{Computation Capacity Constraint}
For each network, the total computation load across all connected vehicles must not exceed the network's available computational capacity. This is represented as:

\begin{equation} \label{eq:comp_res}
\sum_{j=1}^{V} \alpha_{ij} f_{ij} \leq F_i^{\text{max}}, \quad \forall i \in {0, 1, \dots, N}
\end{equation}
where $f_{ij}$ denotes the computational load for vehicle $\nu_j$ connected to network $n_i$, and $F_i^{\text{max}}$ is the maximum computational capacity of network $n_i$.

\subsubsection{Directional Constraint}
Let $L_v$ and $L_n$ be the position vectors of vehicle $\nu_j$ and network $n_i$, respectively. Define the direction vector $\vec{d}_{ij} = L_n - L_v$, and let $\vec{v}j$ be the vehicle’s heading direction. The angular deviation $\theta{ij}$ must satisfy:

\begin{equation} \label{eq:angle}
\theta_{ij} = \cos^{-1}\left( \frac{\vec{d}_{ij} \cdot \vec{v}j}{| \vec{d}{ij} | \cdot | \vec{v}j |} \right) \leq \theta{\text{TH}}
\end{equation}

\subsubsection{Power Constraint}

A vehicle can only connect to a network if the received signal strength (RSS) meets a predefined threshold for reliable communication. This constraint is expressed as:

\begin{equation} \label{eq:power}
P_{ij} \alpha_{ij} \geq P_{\text{TH}}, \quad \forall i, j
\end{equation}
where $P_{ij}$ is the received power from network $n_i$ at vehicle $\nu_j$, and $P_{\text{TH}}$ is the minimum threshold for received signal strength.

\subsubsection{SINR Constraint}

The Signal-to-Interference-plus-Noise Ratio (SINR) must exceed a threshold for reliable communication. This constraint ensures that the SINR at vehicle $\nu_j$ from the network $n_i$ is sufficiently high:

\begin{equation} \label{eq:sinr}
\text{SINR}{ij} \geq \text{SINR}{\text{TH}}, \quad \forall i, j
\end{equation}
where $\text{SINR}_{ij}$ is the SINR at vehicle $\nu_j$ from network $n_i$.

These constraints collectively ensure that the network selection is not only optimal in terms of latency but also feasible under available network and environmental conditions.

\section{Network Selection for V2X Communication}
\label{sec:networkSelection}

\begin{algorithm}[!b]
\caption{Branch-and-Bound Based Optimization Algorithm}
\label{alg:bb}
\begin{algorithmic}[1]  
\STATE \textbf{Given:} Network parameters, application requirements, objective function e.g., Eq.~\eqref{eq:Tj}, and constraint set e.g., Eqs.~\eqref{eq:Tij}--\eqref{eq:Tij_expanded}
\STATE \textbf{Return:} Optimal binary assignment matrix $\boldsymbol{\alpha}^*$
\STATE

\STATE \textbf{Initialization:}
\STATE Flatten the binary matrix $\boldsymbol{\alpha} \in \{0,1\}^{(N+1) \times V}$ into a vector $\mathbf{x} \in \{0,1\}^{(N+1)V}$
\STATE Construct cost vector $\mathbf{f} \in \mathbb{R}^{(N+1)V}$ based on dissemination time $T_{vn}$
\STATE Encode constraints in matrix form: $\mathbf{A}\mathbf{x} \leq \mathbf{b}$, $\mathbf{A}_{\text{eq}}\mathbf{x} = \mathbf{b}_{\text{eq}}$
\STATE Define box bounds: $\mathbf{0} \leq \mathbf{x} \leq \mathbf{1}$
\STATE

\STATE \textbf{Branch-and-Bound Optimization:}
\STATE Relax integer constraints and solve LP: $\min_{\mathbf{x} \in [0,1]^{(N+1)V}} \mathbf{f}^T \mathbf{x}$
\IF{$\mathbf{x}$ is binary}
    \STATE $\mathbf{x}^* \gets \mathbf{x}$
\ELSE
    \STATE Select a fractional component $x_i \in (0,1)$
    \FOR{each branch: $x_i = 0$ and $x_i = 1$}
        \STATE Add constraint $x_i = 0$ or $x_i = 1$
        \STATE Solve the updated subproblem
        \IF{solution is infeasible or worse than current best}
            \STATE Prune this branch
        \ELSE
            \STATE Update best feasible solution $\mathbf{x}^*$
        \ENDIF
    \ENDFOR
    \STATE Repeat until all nodes are either pruned or explored
\ENDIF
\STATE

\STATE \textbf{Final Step:} Reshape optimal vector $\mathbf{x}^*$ into matrix form $\boldsymbol{\alpha}^* \in \{0,1\}^{(N+1) \times V}$
\end{algorithmic}
\end{algorithm}

In this section, we present the network selection mechanism for vehicular communication, specifically targeting V2X applications. The primary goal is to minimize system response time and optimize the network selection for vehicles based on application needs, such as V2X safety or infotainment services. We consider a set of heterogeneous networks, including a vehicular ad hoc network (VANET), 4G, 5G, and other candidate networks, denoted as $\mathcal{N} = \{n_0, n_1, \dots, n_N\}$. Here, $n_0$ represents the current ad hoc network (VANET), while $n_i$ for $i \in \{1, \dots, N\}$ represents the available alternative networks in the vicinity. Each network provides distinct resources (e.g., bandwidth, power, computational capacity), and vehicles in the network are represented as $\mathcal{V} = \{\nu_1, \nu_2, \dots, \nu_V\}$. The decision to connect a vehicle $\nu_j$ to a specific network $n_i$ depends on several factors, such as signal quality (e.g., SNR), available bandwidth, computational resources, and the vehicle’s position relative to the network. These factors are used to optimize the vehicle’s connectivity, reducing the data dissemination time and meeting application requirements like latency and data rate.

\subsection{Network Selection Procedure}

To establish a benchmark for evaluating our proposed adaptive network selection strategy, we modeled the network selection problem as a constrained optimization problem aiming to minimize total data dissemination delay across all vehicles in the V2X environment. This optimization problem incorporates key constraints such as bandwidth availability, computation capacity limits, signal strength thresholds, and directional alignment between vehicles and access points. By formulating this as a Mixed-Integer Linear Programming (MILP) problem, we were able to leverage an exact optimization technique grounded in the branch-and-bound method to determine the optimal network assignment for each vehicle.

The branch-and-bound algorithm in Algorithm 1 explores feasible network selections by branching on binary decision variables and applying linear relaxations to bound subproblems. This ensures globally optimal solutions while minimizing dissemination delay, accounting for both transmission and computation time per vehicle-network pair. Network selection proceeds in multiple steps, allowing vehicles to stay or switch networks based on connectivity, latency, cost, and application type. For example, safety applications prioritize low latency, whereas infotainment may favor higher data rates, as illustrated in Algorithm 2.

\begin{algorithm}[H]
\caption{Proposed Adaptive Network Selection Based on Latency and Application Type}
\label{alg:heuristic}
\begin{algorithmic}[1]
\REQUIRE Set of vehicles $\mathcal{V} = \{\nu_1, \nu_2, \dots, \nu_V\}$ \\
\hspace{\algorithmicindent}Set of networks $\mathcal{N} = \{n_1, n_2, \dots, n_N\}$ \\
\hspace{\algorithmicindent}Application requirements and current network assignment for each vehicle
\ENSURE Network selection vector $\boldsymbol{\alpha} = [\alpha_{ij}]_{N \times V}$

\FOR{each vehicle $\nu_j \in \mathcal{V}$}
    \STATE $\text{best\_metric} \gets \infty$ (if $\text{App}_j$ is safety) or $0$ (otherwise)
    \STATE $\text{selected\_network} \gets \emptyset$
    
    \FOR{each network $n_i \in \mathcal{N}$}
        \IF{link $(i,j)$ is available}
            \STATE Transmission delay:
            \[
                T_{ij}^{\text{trans}} = \frac{s_j}{r_{ij}} + \Delta_i + \frac{P_{ij}^{\text{idle}} \cdot \sigma + P_{ij}^c \cdot T_{ij}^c}{P_{ij}^s}
            \]
            
            \STATE Computation delay:
                $T_{ij}^{\text{comp}} = \frac{c_j}{f_{ij}}$

            \STATE Handover delay:
            \IF{$n_i$ is different from current network of $\nu_j$}
                \STATE Sample $T_{ij}^{\text{handover}} \sim \mathcal{N}(\mu\,\text{ms},\,\sigma^2\,\text{ms}^2)$
            \ELSE
                \STATE $T_{ij}^{\text{handover}} \gets 0$
            \ENDIF
            
            \STATE Total delay:
                $T_{ij} = T_{ij}^{\text{trans}} + T_{ij}^{\text{comp}} + T_{ij}^{\text{handover}}$
            
            \STATE Direction alignment:
                $D_{ij} = \frac{\mathbf{d}_j - \mathbf{p}_j}{\|\mathbf{d}_j - \mathbf{p}_j\|} \cdot \frac{\mathbf{p}_i - \mathbf{p}_j}{\|\mathbf{p}_i - \mathbf{p}_j\|}$

            \IF{$\text{App}_j$ is safety-critical}
                \STATE $\text{score}_{ij} \gets T_{ij} - \lambda \cdot D_{ij}$
                \IF{$\text{score}_{ij} < \text{best\_metric}$}
                    \STATE $\text{best\_metric} \gets \text{score}_{ij}$
                    \STATE $\text{selected\_network} \gets i$
                \ENDIF
            \ELSIF{$\text{App}_j$ is infotainment}
                \STATE $\text{score}_{ij} \gets -r_{ij} + \lambda \cdot (1 - D_{ij})$
                \IF{$\text{score}_{ij} > \text{best\_metric}$}
                    \STATE $\text{best\_metric} \gets \text{score}_{ij}$
                    \STATE $\text{selected\_network} \gets i$
                \ENDIF
            \ENDIF

        \ENDIF
    \ENDFOR

    \IF{$\text{selected\_network} \neq \emptyset$}
        \STATE $\alpha_{ij} \gets 1$ \COMMENT{Assign vehicle $j$ to network $i$}
    \ENDIF
\ENDFOR

\STATE \textbf{Post-Selection Resource Allocation:}
\FOR{each network $n_i \in \mathcal{N}$}
    \STATE $B_i^{\text{used}} \gets \sum_{j=1}^V \alpha_{ij} \cdot b_{ij}$, 
    $F_i^{\text{used}} \gets \sum_{j=1}^V \alpha_{ij} \cdot f_{ij}$

    \IF{$B_i^{\text{used}} > B_i^{\text{max}}$ OR $F_i^{\text{used}} > F_i^{\text{max}}$}
        \STATE Sort $\{\nu_j \mid \alpha_{ij} = 1\}$ by priority (safety first)
        \WHILE{resources exceeded}
            \STATE Remove lowest priority vehicle: $\alpha_{ij} \gets 0$
            \STATE Recompute $B_i^{\text{used}}, F_i^{\text{used}}$
        \ENDWHILE
    \ENDIF
\ENDFOR

\RETURN $\boldsymbol{\alpha}$
\end{algorithmic}
\end{algorithm}

\subsection{ANS-V2X Algorithm Description}

This optimal solution provides an upper performance bound as a benchmark for assessing the proposed ANS-V2X scheme. By comparing the adaptive, application-aware selection strategy against this optimal benchmark, ANS-V2X achieves near-optimal performance in practical scenarios with varying vehicle densities and heterogeneous network conditions while maintaining computational efficiency suitable for real-time deployment. The proposed algorithm is designed to perform application-aware network selection for vehicles in a V2X communication environment. Initially, all vehicles are connected through DSRC. The algorithm considers heterogeneous wireless networks and selects the most suitable one whenever the next-hop vehicle is not available. \textcolor{red}{The proposed ANS-V2X framework supports multiple V2X communication modes, including V2V, V2I, and V2N. Based on real-time network conditions, the algorithm selects either a neighboring vehicle (for direct V2V transmission) or an access network (e.g., LTE, 5G) for infrastructure-based communication. This selection is guided by a unified utility function that incorporates SINR, latency sensitivity, and queue delay, ensuring compatibility with diverse V2X applications such as safety-critical messaging, infotainment, and traffic updates.}

For each vehicle \( \nu_j \), the algorithm begins by identifying all available networks \( n_i \in \mathcal{N} \). For each network, it measures the received signal power \( P_{ij} \) and the noise level \( N_{ij} \), then calculates the SINR as \( \text{SINR}_{ij} = \frac{P_{ij}}{N_{ij}} \). It also computes the directional angle \( \theta_{ij} \) between the vehicle and the network access point to ensure that the network lies within the vehicle’s forward direction. A network is considered a valid candidate only if it satisfies the received signal strength \( P_{ij} \) exceeds a predefined threshold \( P_{\text{TH}} \), the SINR is greater than a threshold \( \text{SINR}_{\text{TH}} \), and the direction angle \( \theta_{ij} \) is less than or equal to \( 180^\circ \).

For all valid networks, the algorithm computes the \textit{transmission delay} \( T_{ij}^{\text{trans}} \) and the \textit{computation delay} \( T_{ij}^{\text{comp}} \), defined as:
\[
T_{ij}^{\text{trans}} = \frac{s_j}{r_{ij}} + \Delta_i + \frac{P_{ij}^{\text{idle}} \cdot \sigma + P_{ij}^c \cdot T_{ij}^c}{P_{ij}^s}, \quad
T_{ij}^{\text{comp}} = \frac{c_j}{f_{ij}}
\]
\textcolor{red}{The total dissemination time \( T_{ij} \) is the sum of transmission delay, computation delay, and a switching delay term \( T_{ij}^{\text{handover}} \), which accounts for handover latency when a vehicle transitions between networks.}

Depending on the type of application required by the vehicle, the network selection strategy changes. If the vehicle requires a safety application, the algorithm selects the network that minimizes the total dissemination delay \( T_{ij} \). In contrast, if the vehicle is running an infotainment application, it selects the network that offers the maximum data rate \( r_{ij} \).

After identifying the best network according to the application type, the algorithm checks whether the \textit{bandwidth} and \textit{computation capacity constraints} are satisfied. Specifically, it ensures that the cumulative bandwidth and computation loads from all connected vehicles do not exceed the total bandwidth \( B_i \) or computation capacity \( F_i^{\text{max}} \) of the network. If the constraints are met, the vehicle is handed over to the selected network. If not, the algorithm iterates over the remaining networks in the ranked candidate list until a feasible one is found. This process repeats for all vehicles in the system, resulting in an optimized network selection vector \( \boldsymbol{\alpha} \) that minimizes overall data dissemination time while ensuring application-aware connectivity and maintaining system-wide constraints.

\section{Time Complexity Analysis}

The time complexity of the proposed network selection algorithm is primarily governed by the number of vehicles in the system and the number of available networks each vehicle must evaluate. For every vehicle, the process begins with the discovery and assessment of all accessible networks. This involves computing performance metrics such as signal-to-noise ratio, transmission delay, or achievable data rate, depending on the application's nature. Since each vehicle evaluates all \( N \) networks, this results in a per-vehicle computational effort of \( O(N) \). During the application-specific evaluation, vehicles apply additional logic to prioritize metrics based on whether the application is latency-sensitive (e.g., safety-critical) or throughput-driven (e.g., infotainment). This evaluation phase also adheres to a complexity of \( O(N) \). If the algorithm employs sorting to rank the candidate networks based on the computed scores, the per-vehicle complexity increases to \( O(N \log N) \). However, in scenarios where the best candidate can be determined through a single pass, the complexity remains linear. Consequently, when accounting for all \( V \) vehicles in the system, the total computational complexity becomes \( O(V \times N) \) in the linear case or \( O(V \times N \log N) \) if sorting is applied. \textcolor{red}{In contrast, the worst-case complexity of Algorithm-1 is \( O(N^2) \), which arises when evaluating all network-vehicle combinations for scoring and feasibility. Despite this, the overall runtime remains acceptable for real-time scenarios due to the algorithm’s lightweight operations and localized decision-making.}

\textcolor{red}{For comparison, the Q-learning-based method incurs higher computational overhead due to iterative learning. Each vehicle maintains a Q-table of size \( S \times N \), where \( S \) is the number of discrete states. Over \( E \) episodes, the complexity becomes \( O(V \times E \times N) \), considering action-value updates, exploration, and reward computations. While this approach enables learning optimal policies over time, it is less suitable for real-time operation without pretraining.}

\textcolor{red}{On the other hand, the MILP-V2X model involves solving a combinatorial optimization problem with \( N \times V \) binary decision variables and multiple linear constraints. Using the Branch-and-Bound method, its worst-case complexity grows exponentially, making it computationally infeasible for large-scale or time-sensitive deployments.}

This complexity reflects the algorithm’s scalable design, enabling each vehicle to dynamically select the most suitable network from multiple heterogeneous options according to its application-specific demands.

The network selection algorithm aims to select the optimal network for each vehicle, considering application-specific demands and multiple constraints. The goal is to minimize latency and maximize throughput while adhering to various network resource constraints, such as bandwidth, computation power, signal strength, and directional constraints.

\subsection{Simulation Setup}

The simulation setup evaluates the performance of two algorithms designed to optimize network selection in V2X systems. The first is the MILP-V2X algorithm, which represents the optimal network selection approach using the Branch-and-Bound method. This algorithm solves the mixed-integer linear programming (MILP) formulation to achieve the optimal network assignment that minimizes data dissemination time while respecting constraints such as bandwidth, computational resources, and network direction. \textcolor{red}{Vehicle mobility is modeled using the Simulation of Urban MObility (SUMO) framework with realistic urban road topologies, as described in \cite{lopez2018microscopic}. To capture the dynamics of vehicular movement, each vehicle’s position and velocity are updated at 100 ms intervals throughout the simulation. These updates influence signal quality (e.g., SINR), connectivity, and queue delay due to variable distances from access points and potential handovers between networks. Additionally, ANS-V2X is designed with computational simplicity to meet the real-time processing and memory limitations typical of On-Board Units (OBUs). The algorithm avoids iterative solvers or deep models, and all computations are completed within 15 ms on standard embedded hardware, making it practical for deployment in real vehicular systems with limited CPU and energy budgets.}

\textcolor{red}{The impact of mobility is explicitly integrated into the network selection process. As vehicles move across coverage zones of different Radio Access Technologies (RATs), the candidate network set and its associated QoS parameters dynamically change. This leads to adaptive re-evaluation of utility values, allowing ANS-V2X and other schemes to make informed decisions based on time-varying link quality and application needs.}

\textcolor{red}{Mobility-induced transitions between networks are modeled by introducing a handover delay term \( T_{ij}^{\text{handover}} \), applied only when a vehicle switches between different networks. This delay is sampled from a Gaussian distribution with a mean of 20~ms and a standard deviation of 5~ms, reflecting real-world inter-RAT switching latencies (e.g., LTE-to-5G), as suggested in~\cite{mahmood2019handover}. The same mobility-aware handover model is uniformly applied across all three schemes to ensure fair evaluation under dynamic vehicular conditions.}


\begin{table}[!t]
\footnotesize
\centering
\caption{Simulation Parameters for MILP-V2X, Q-Learning, and ANS-V2X Algorithms}
\label{tab:simparams}
\renewcommand{\arraystretch}{1.2}
\begin{tabular}{|p{3.4cm}|p{4.8cm}|}
\hline
\textbf{Parameter} & \textbf{Value/Description} \\ \hline
Network Types & \textcolor{red}{4G LTE, 5G NR, IEEE 802.11p (DSRC)} \\ \hline
Vehicular Density & \textcolor{red}{Low: 100, Medium: 200, High: 300 vehicles/km$^2$} \\ \hline
Application Types & Safety (e.g., collision avoidance), Infotainment (e.g., video streaming) \\ \hline
Bandwidth per Network & \textcolor{red}{4G: 20 MHz, 5G: 100 MHz, 802.11p: 10 MHz} \\ \hline
SNR Threshold & \textcolor{red}{Minimum 15 dB for reliable connectivity} \\ \hline
Latency Requirements & \textcolor{red}{Safety: $\leq$ 50 ms, Infotainment: $\leq$ 100 ms} \\ \hline
Computational Capacity & \textcolor{red}{4G: 1 GHz, 5G: 2 GHz, 802.11p: 500 MHz (per OBU)} \\ \hline
Power Model & \textcolor{red}{Idle: 10 W, Active: 30 W (per vehicle)} \\ \hline
\textcolor{red}{Mobility Model} & \textcolor{red}{SUMO-based urban mobility with realistic vehicle trajectories~\cite{lopez2018microscopic}} \\ \hline
\textcolor{red}{Propagation Model} & \textcolor{red}{Log-distance path loss with urban shadowing} \\ \hline
Simulation Duration & \textcolor{red}{500 seconds per run with multiple Monte Carlo iterations} \\ \hline
Comparison Metrics & \textcolor{red}{Total Utility, E2E Latency, Network Load Distribution, Convergence Speed} \\ \hline
\end{tabular}
\end{table}

ANS-V2X is a heuristic-based approach designed for adaptive network selection. This algorithm selects the most appropriate network based on real-time latency measurements and the specific application requirements of each vehicle. It dynamically evaluates the network conditions, such as signal strength, latency, and available bandwidth, and assigns the optimal network considering application priorities. The ANS-V2X algorithm offers a balance between computational efficiency and network performance, making it suitable for real-time vehicular environments.

The algorithms are implemented and compared in terms of their performance under varying vehicular densities and network conditions. The key metrics for comparison include data dissemination time, computational load, and overall system throughput. Simulations are conducted across different network scenarios (e.g., 4G, 5G, ad hoc) to assess the adaptability and robustness of each approach. \textcolor{red}{Furthermore, real-world signal disruptions and coverage holes are not explicitly modeled in this work, as the focus is on network selection under varying densities.}

Table~\ref{tab:simparams} provides an overview of the key parameters used in the simulation setup for the ANS-V2X, MILP-V2X, and Q-learning algorithms. These parameters are chosen to reflect realistic vehicular communication scenarios and enable a fair and comprehensive comparison across all three approaches.

\begin{table*}[hbt]
\centering
\caption{Qualitative Comparison of Network Selection Methods}
\begin{tabular}{|l|c|c|c|c|c|c|}
\hline
\textbf{Method} & \textbf{Optimality} & \textbf{Runtime} & \textbf{Scalability} & \textbf{Training} & \textbf{Real-time} & \textbf{Deployable} \\
\hline
MILP-V2X & Global & High ($>$100 ms) & Poor & No & No & No \\
Q-learning & Near-optimal & Moderate (20--40 ms) & Moderate & Yes & Partial & Limited \\
ANS-V2X & Near-optimal (5--10\%) & Low ($\leq$15 ms) & Good & No & Yes & Yes \\
\hline
\end{tabular}
\label{tab:runtime_comparison}
\end{table*}

\section{Results and Discussion}

\textcolor{red}{This section presents a detailed performance evaluation of the proposed ANS-V2X algorithm in comparison with two benchmark approaches: the optimization-based MILP-V2X scheme and a machine learning-based method using a Q-learning agent \cite{cheng2022qlearningv2x}. The Q-learning agent learns optimal network selection actions based on observed SINR, delay, and utility by maintaining a Q-table indexed by network states and updating it through feedback over multiple episodes. In real-world deployments, SINR, delay, and position estimates are subject to estimation errors caused by noise. ANS-V2X demonstrates inherent robustness to such errors due to its threshold-driven and localized decision logic. Small inaccuracies in SINR or delay estimates are mitigated through periodic reassessment, while minor GPS noise in vehicle positioning has a limited impact on directionality computation \( D_{ij} \). Unlike MILP-V2X, which relies on globally precise input, and Q-learning, which is sensitive to reward feedback, ANS-V2X avoids instability by preventing frequent switching unless utility gains are significant.}

\textcolor{red}{Table~\ref{tab:runtime_comparison} summarizes the qualitative runtime behavior and practical deployment readiness of the three evaluated approaches. While MILP-V2X offers global optimality, it is computationally expensive and unsuitable for real-time use. Q-learning adapts over time but requires extensive training. In contrast, ANS-V2X achieves near-optimal results with significantly lower runtime, making it ideal for deployment in latency-sensitive V2X systems. The simulations are conducted under diverse scenarios by varying the number of available networks ($N = 1$ to $5$) and the number of vehicles ($V = 5$, $9$, and $12$), reflecting realistic conditions encountered in dynamic V2X environments. Each network is characterized by distinct QoS attributes such as bandwidth, latency, and reliability, while each vehicle is associated with either a safety-critical or infotainment-based application demand. This evaluation framework highlights the adaptability, scalability, and effectiveness of the proposed ANS-V2X approach in handling application-aware network selection under heterogeneous and latency-sensitive communication requirements.}

\begin{figure}[!b]
  \centering
  \includegraphics[width=0.46\textwidth]{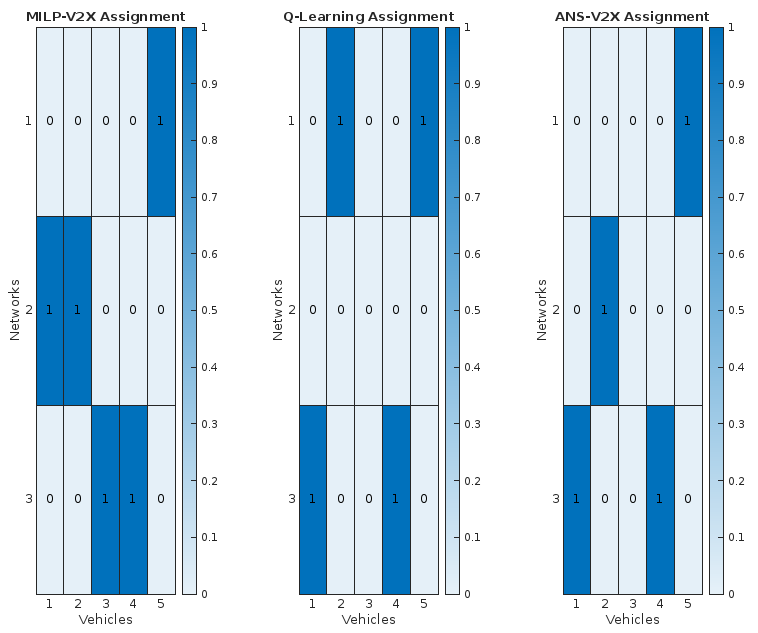}
  \caption{Vehicle-to-Network assignment heatmap for \(N = 3\) networks and \(V = 5\) vehicles. MILP-V2X achieves a globally optimal and balanced assignment but with higher computational complexity.}
  \label{fig:fig1}
\end{figure}

\textcolor{red}{The MILP-V2X approach employs a branch-and-bound optimization framework that guarantees globally optimal assignments across all scenarios. It serves as the theoretical upper bound for achievable utility and minimal latency. However, due to the exponential growth in complexity with increasing problem size, MILP-V2X exhibits substantial computational latency. For example, in medium-density scenarios (e.g., 50 vehicles), it often exceeds 100\,ms in computation time, making it impractical for real-time applications. The Q-learning-based approach introduces a reinforcement learning framework where each vehicle learns optimal network selection strategies over repeated interactions with the environment. This method offers adaptability to dynamic network conditions and can discover effective policies without prior knowledge of system parameters. However, its performance depends on careful tuning of learning parameters and sufficient training iterations. While Q-learning can reach near-optimal solutions in some configurations, its convergence speed and occasional inconsistency make it less suitable for latency-critical V2X systems.}

\textcolor{red}{The proposed ANS-V2X heuristic provides a lightweight and practical alternative by incorporating application-aware logic with real-time feasibility. It approximates the MILP-V2X performance while reducing computation time significantly. In the same medium-density settings, ANS-V2X consistently completes decisions in under 15\,ms, making it well-suited for real-time vehicular environments. Its structured prioritization of vehicle demands, combined with adaptive resource allocation, enables effective network-to-vehicle assignments even under stringent latency and QoS constraints.}

\begin{figure}[!b]
  \centering
  \includegraphics[width=0.46\textwidth]{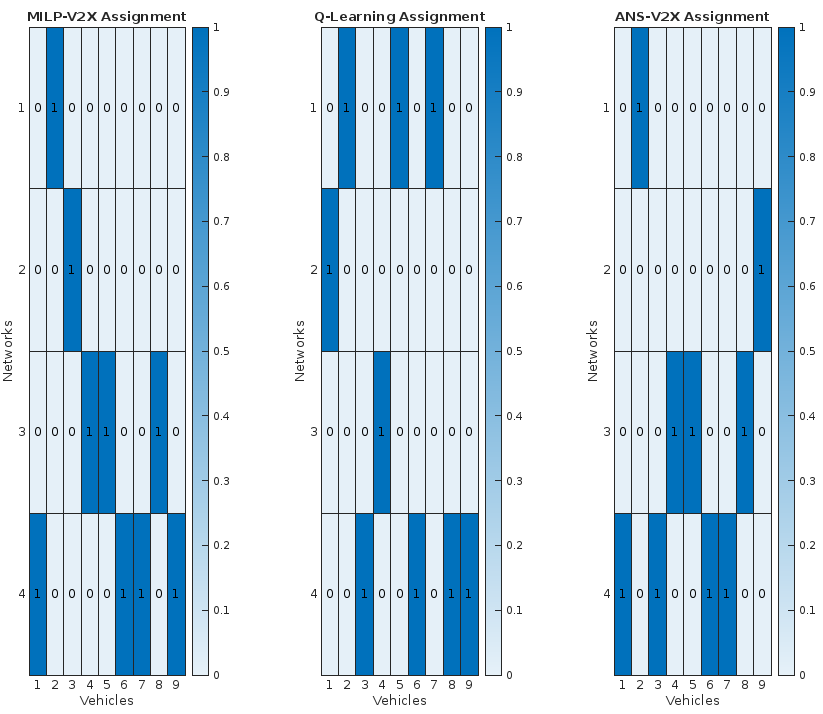}
  \caption{Vehicle-to-Network assignment heatmap for \(N = 4\) networks and \(V = 9\) vehicles. Q-Learning and ANS-V2X show distinct patterns with ANS-V2X favoring latency-aware selections.}
  \label{fig:fig4}
\end{figure}

\begin{figure}[!t]
  \centering
  \rotatebox{90}{%
    \includegraphics[width=0.64\textwidth]{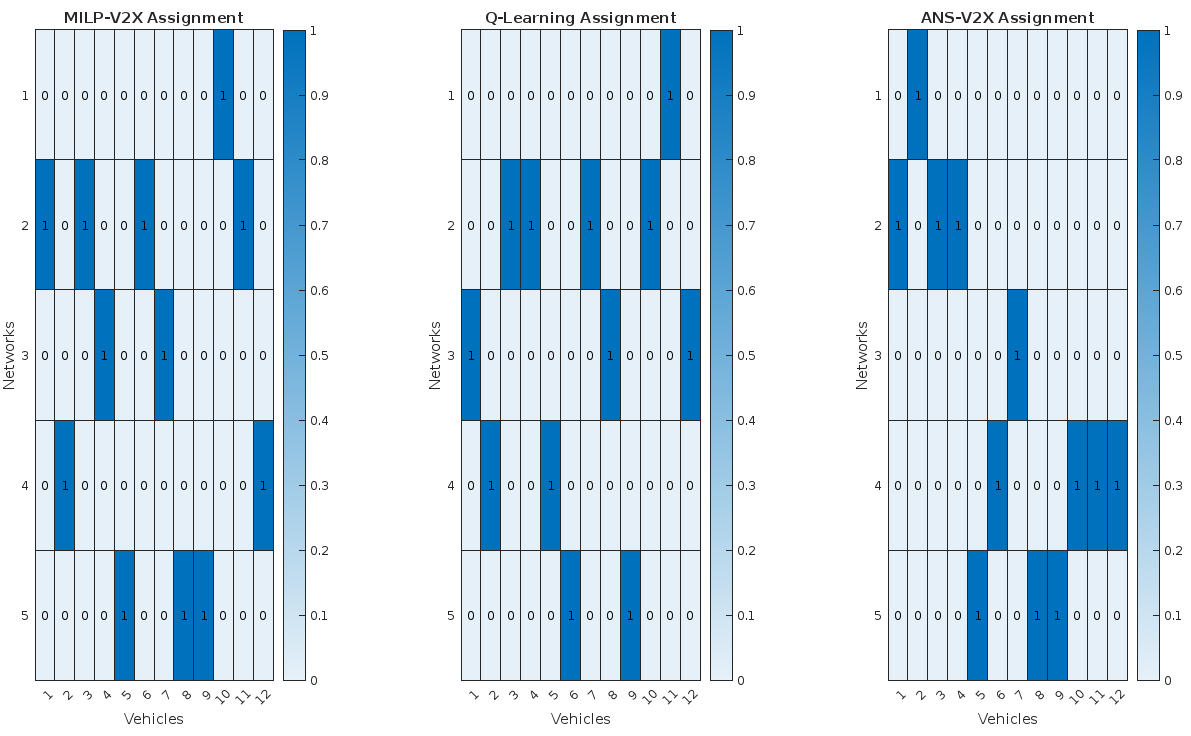}
  }
  \caption{Vehicle-to-network assignment heatmap for \(N = 5\) networks and \(V = 12\) vehicles. Despite the increased complexity, ANS-V2X achieves a near-optimal assignment while maintaining computational efficiency.}

  \label{fig:fig11}
\end{figure}

\textcolor{red}{Figures~\ref{fig:fig1}, \ref{fig:fig4}, and \ref{fig:fig11} present heatmaps that illustrate both the achieved utility and the network-to-vehicle assignment patterns for ANS-V2X, MILP-V2X, and Q-learning under varying network and vehicle densities. The x-axis represents the number of network nodes (e.g., RSUs), and the y-axis denotes the number of vehicles. Color intensity reflects the normalized utility, with warmer shades indicating better performance. The configurations shown, $(N = 3, V = 5)$, $(N = 4, V = 9)$, and $(N = 5, V = 12)$, demonstrate how each scheme allocates communication resources as system complexity increases. ANS-V2X consistently maintains high utility across diverse density conditions, showcasing adaptability and scalability. MILP-V2X, while producing near-optimal assignments that distribute load evenly, becomes less practical in dense settings due to computational limitations. Q-learning shows inconsistent patterns, highlighting slower convergence and less effective adaptation under dynamic conditions. This behavior is particularly evident in Figure~\ref{fig:fig1}, where a balanced mapping ensures minimal contention and efficient exploitation of network resources. However, as the number of networks and vehicles increases, as seen in Figure~\ref{fig:fig11}, the computational burden of MILP-V2X becomes prohibitive due to the exponential search space explored by the branch-and-bound algorithm. This scalability challenge limits its applicability in real-time V2X scenarios where rapid decision-making is essential.}

\textcolor{red}{Q-Learning offers a learning-based alternative that adapts over time through interaction with the environment. While it performs competitively in smaller configurations (e.g., Figure~\ref{fig:fig1}), its effectiveness degrades in more complex settings such as Figure~\ref{fig:fig11}. This degradation stems from several factors: (i) convergence requires a large number of episodes, which may be infeasible in real-time vehicular environments; (ii) limited exploration of the large state-action space can lead to suboptimal policy learning; and (iii) resource constraints and coverage prohibitions further restrict the set of feasible actions. As a result, the learned policies may assign multiple vehicles to overloaded or suboptimal networks, negatively impacting utility and latency.}

\textcolor{red}{The proposed ANS-V2X heuristic demonstrates robust performance across all configurations, with heatmaps showing context-aware and latency-sensitive assignments. In Figure~\ref{fig:fig4}, for instance, ANS-V2X intelligently prioritizes low-latency networks for delay-sensitive applications while maintaining feasible resource allocation. Despite not guaranteeing global optimality, ANS-V2X remains consistently close to the MILP benchmark while operating at significantly lower computational cost. This trade-off makes it highly suitable for deployment in dynamic and time-constrained V2X environments. To reflect practical deployment, we acknowledge the presence of handover latency when switching between RATs. In our framework, handover delays are assumed to follow a Gaussian distribution with a mean of 20\,ms, based on typical LTE-to-5G inter-RAT switching delays~\cite{mahmood2019handover}.}

\begin{figure}[!t]
  \centering
  \includegraphics[width=0.48\textwidth]{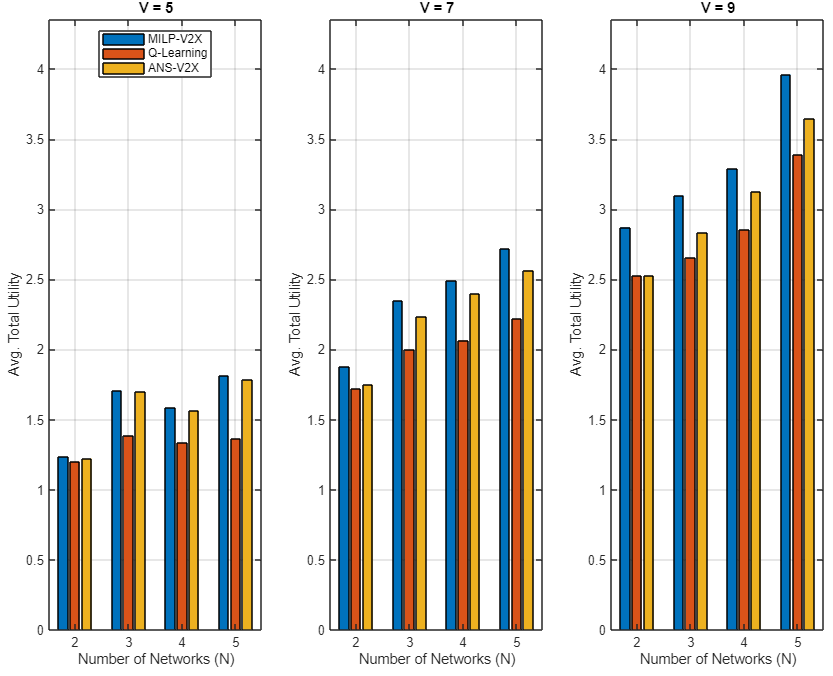}
  \caption{Average total utility versus number of networks ($N=2$ to $N=5$) for vehicle counts $V=5$, $V=7$, and $V=9$ across MILP-V2X, Q-Learning, and ANS-V2X.}
  \label{fig:fig13}
\end{figure}

\begin{figure}[!t]
  \centering
  \includegraphics[width=0.48\textwidth]{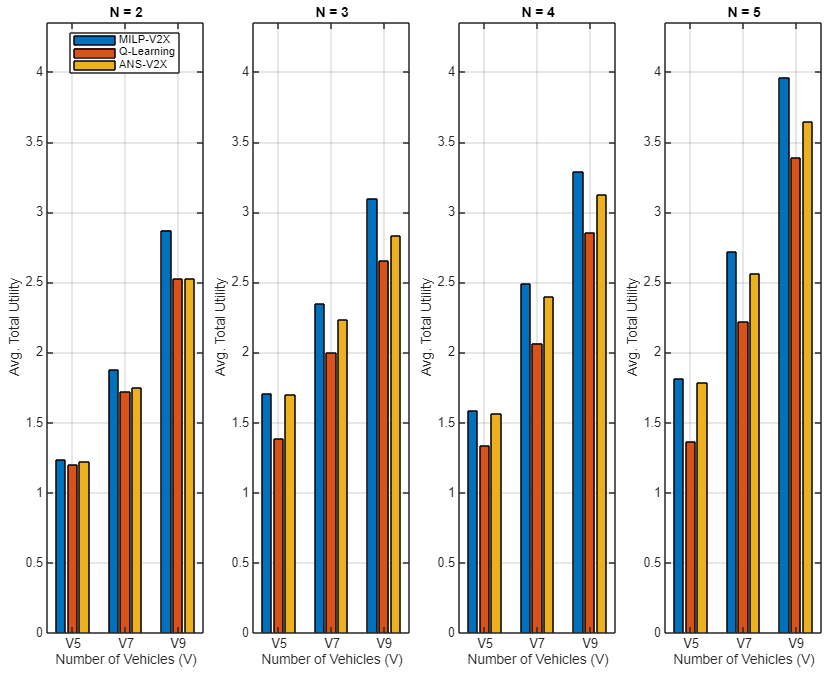}
  \caption{Average total utility versus number of vehicles ($V=5$, $V=7$, and $V=9$) for network configurations ($N=2$ to $N=5$) across MILP-V2X, Q-Learning, and ANS-V2X.}
  \label{fig:fig14}
\end{figure}

\textcolor{red}{Figures \ref{fig:fig13} and \ref{fig:fig14} present a comparative evaluation of average total utility under varying network and vehicle configurations. As expected, MILP-V2X consistently delivers the highest average utility. This is due to its global optimization mechanism that exhaustively explores all feasible assignments via branch-and-bound, ensuring maximal utility aggregation across all vehicles and networks. Q-Learning, being a reinforcement learning-based strategy, performs relatively well but shows degradation in highly dynamic configurations. This is attributed to its need for sufficient exploration during training and the stochastic nature of its learning process. As the environment scales with higher vehicle and network counts, Q-Learning struggles to converge to consistently optimal policies, particularly in resource-constrained conditions. This leads to lower average utility when compared to MILP and ANS.}

\begin{figure}[!b]
  \centering
  \includegraphics[width=0.46\textwidth]{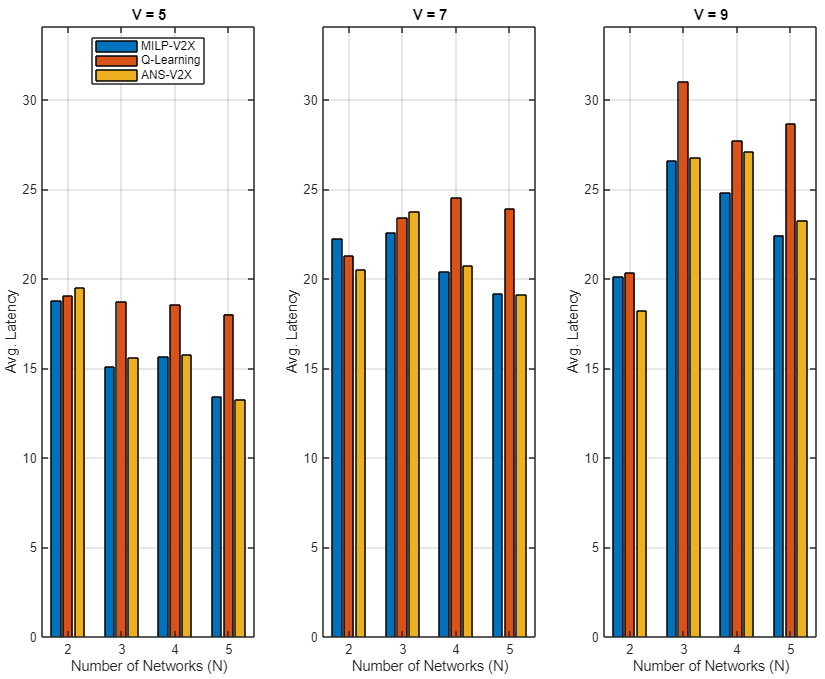}
  \caption{Average latency versus number of networks ($N=2$ to $N=5$) for vehicle counts $V=5$, $V=7$, and $V=9$ across MILP-V2X, Q-Learning, and ANS-V2X.}
  \label{fig:fig15}
\end{figure}

\begin{figure}[!t]
  \centering
  \includegraphics[width=0.48\textwidth]{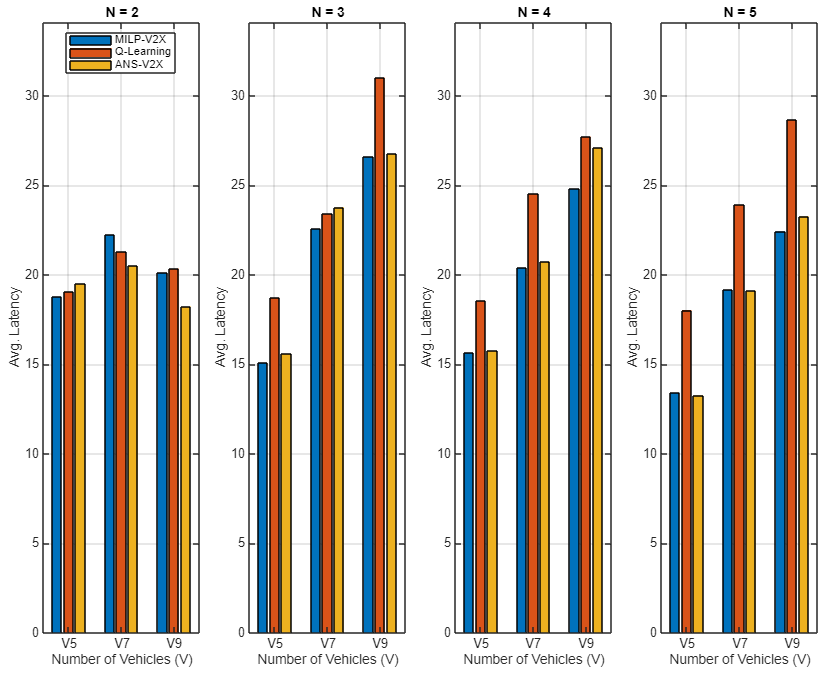}
  \caption{Average latency versus number of vehicles ($V=5$, $V=7$, and $V=9$) across different network counts ($N=2$ to $N=5$) for MILP-V2X, Q-Learning, and ANS-V2X.}
  \label{fig:fig16}
\end{figure}

\textcolor{red}{The proposed ANS-V2X demonstrates competitive performance, closely trailing the MILP-V2X in utility across most scenarios. Unlike MILP, which incurs a significant computational burden, ANS-V2X utilizes a latency-aware heuristic that enables fast decision-making. Notably, under tighter resource constraints or increasing vehicle density, ANS-V2X even surpasses Q-Learning by better aligning network selection with application-specific latency and resource demands. This makes ANS-V2X particularly attractive for real-time vehicular environments, where quick decisions must be made with limited computational overhead.}

\textcolor{red}{Figure \ref{fig:fig14} reinforces this by illustrating that while MILP-V2X achieves theoretical optimality, the gap between it and ANS-V2X remains small across growing vehicle densities. Conversely, Q-Learning shows performance volatility as the action space expands. These results highlight the scalability and robustness of ANS-V2X as a near-optimal solution with practical advantages over both complex optimization (MILP) and learning-based (Q-Learning) alternatives.}

\textcolor{red}{Figures \ref{fig:fig15} and \ref{fig:fig16} shift focus to average latency, a critical metric for V2X applications, particularly for safety-critical data transmission. Across all settings, ANS-V2X consistently yields the lowest latency, thanks to its design that integrates latency sensitivity directly into the assignment logic. By dynamically mapping delay-sensitive tasks to lower-latency networks, ANS-V2X maintains responsiveness under both low and high vehicle loads. While MILP-V2X ensures balanced utility, its latency performance often suffers since its objective function is not explicitly delay-aware. Although MILP does achieve occasional latency minimization indirectly, it does not prioritize delay as a first-class metric. Q-Learning, meanwhile, exhibits variable latency performance. Its decentralized decision-making and limited training episodes lead to inconsistent results, especially under denser configurations where convergence takes longer.}

\textcolor{red}{The performance gap is most notable in Figure \ref{fig:fig16}, where ANS-V2X continues to outperform as vehicle count increases. The only minor deviation is observed when network diversity is limited and vehicle count is high (e.g., $N=2$, $V=9$), where all schemes approach similar latency due to network saturation. Still, ANS-V2X maintains a marginal advantage even under these constraints. MILP-V2X provides a theoretical upper bound on utility but lacks real-time feasibility. Q-Learning is adaptive but suffers under complexity and convergence limitations. ANS-V2X offers a compelling middle ground, efficient, latency-aware, and scalable, making it the most suitable candidate for deployment in practical V2X systems.}

\section{Conclusion}
\label{07_Conclusion}

In this work, a heuristic-based adaptive network selection framework, ANS-V2X, was proposed to meet the latency-sensitive and application-aware demands of V2X systems under dynamic network and vehicle densities. The algorithm selects suitable networks by evaluating real-time latency, computation load, and application type. It was compared against MILP-V2X, an optimal but slow Branch-and-Bound solver, and Q-learning, a reinforcement learning method with slower convergence in dense environments. While MILP-V2X showed slightly better utility in sparse settings, its high computation time limited real-time use.
\textcolor{red}{ANS-V2X consistently achieved near-optimal utility (within 5–10\% of MILP) with much lower latency and sub-15 ms decision time, without requiring pretraining or centralized solvers. This efficiency comes from its lightweight, context-aware logic that adapts quickly to network and mobility changes.}
A runtime comparison confirmed that ANS-V2X remained stable and fast, while MILP-V2X and Q-learning incurred higher overhead.
\textcolor{red}{This makes ANS-V2X the most practical among the three, balancing decision quality with runtime efficiency for real-time deployment. Future work includes integrating deep RL, predictive mobility models, and real-world testing to improve scalability and robustness.}




\begin{IEEEbiography}
[{\includegraphics[width=1in,height=1.25in,clip,keepaspectratio]{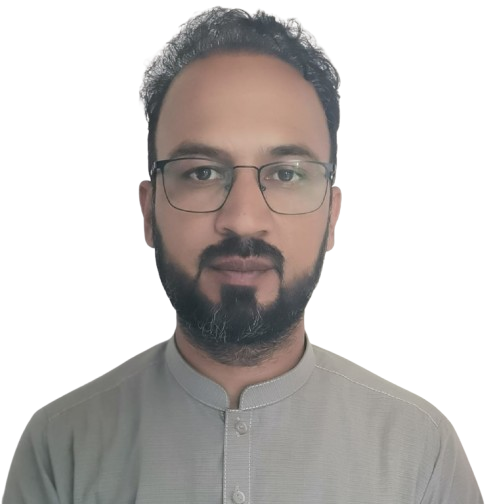}}]
   {Muhammad Zia Ul Haq}  is a PhD scholar in the Electrical Engineering department at COMSATS University Islamabad (Wah Campus), where he is pursuing his research under the prestigious HEC Indigenous-5000 PhD Scholarship. He holds a Master's degree in Computer Science from the University of Malakand (2018) and a Bachelor's in Telecommunication from Hazara University Mansehra. His research focuses on Wireless Sensor Networks (WSNs), Heterogeneous Networks, and V2X Communications.
\end{IEEEbiography}


\begin{IEEEbiography}
[{\includegraphics[width=1in,height=1.25in,clip,keepaspectratio]{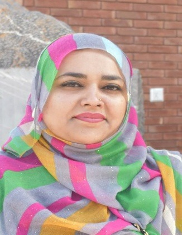}}]
   {Nadia Nawaz Qadri} (Senior Member, IEEE) is currently working as a Professor and Chairperson of the Computer Engineering Department at COMSATS University Islamabad. She completed her B.E. (Computer Systems Engineering) and M.E (Communication Systems and Networks) from Mehran University of Engineering and Technology, Jamshoro in Dec. 2002 and Dec 2004 respectively. She served as a lab lecturer at the same university from 2003 to 2004, and then she worked as a lecturer at Fatima Jinnah Women's University, Rawalpindi, from 2004 to 2005. She joined the COMSATS Institute of Information Technology (CIIT), Wah Campus in Oct 2005. From there, she proceeded to UK for doctorate in Electronic Systems Engineering funded by CIIT, under HEC’s "Faculty Development Program" in 2006. She finished her PhD in June 2010 and joined her parent department as Assistant Professor. She has more than 21 years of teaching and research experience at renowned universities in Pakistan. Her research interests include video streaming, mobile ad hoc networks and vehicular ad hoc networks, P2P Networks, wireless sensor networks, Video and Image Processing, 4G and 5G Communication, Smart Grid, Internet of Things, and Multicore Reconfigurable Architectures. 
\end{IEEEbiography}


\begin{IEEEbiography}
[{\includegraphics[width=1in,height=1.25in,clip,keepaspectratio]{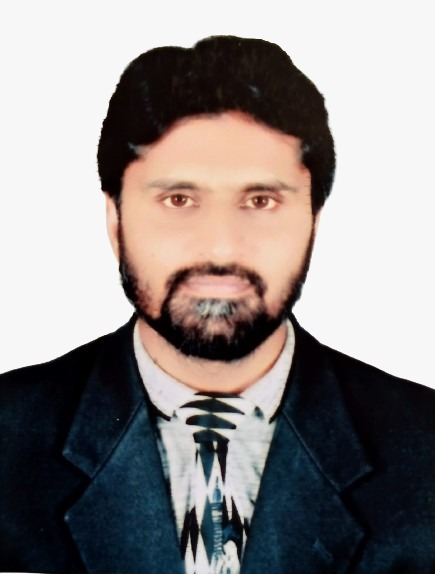}}]
{Omer Chughtai} (Member, IEEE) is a Tenured Associate Professor at COMSATS University Islamabad, Wah Campus, with over 17 years of experience in computer science and engineering. He holds a B.Eng. from UET Taxila, an MS from COMSATS, and a Ph.D. from Universiti Teknologi PETRONAS, Malaysia (fully funded). His research focuses on IoT networks, AI, cross-layer protocol design, and vehicular/flying ad hoc networks. He has published widely, received the Research Productivity Award, secured research grants, and holds a patent. Dr. Chughtai has served in key roles such as Chair of IEEE CCNC 2025 and Event Chair of FIT’24. He leads research in heterogeneous wireless networks, with emphasis on V2X communication, disaster response, and resource optimization.

\end{IEEEbiography}


\begin{IEEEbiography}[{\includegraphics[width=1in,height=1.25in,clip,keepaspectratio]{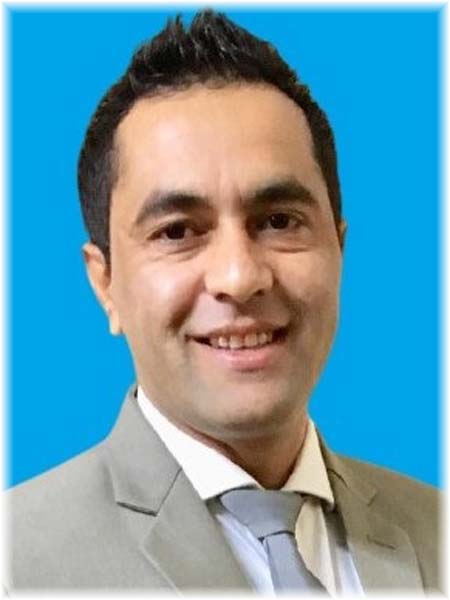}}]{SADIQ AHMAD} (M'19) is currently serving as an assistant professor in the Department of
Electrical Engineering, COMSATS University Islamabad, Wah Cantt., Pakistan. He obtained his MS and Ph.D. degrees, both in Power Engineering from COMSATS University Islamabad, Wah Cantt, in 2014 and 2020, respectively. He obtained his B.Sc. in Electrical Engineering from the University of Engineering and Technology, Peshawar, Pakistan in 2009. He served in Water and Power Development Authority (WAPDA) Pakistan for one year and then joined Etisalat Telecommunication Company until April 2012. He won the Best Research Paper award (2015) under the patronage of the Higher Education Commission, Government of Pakistan. His research interests include privacy preservation in smart meters, blockchain, security and privacy issues in smart grids, cognitive radio networks, energy efficiency, green energy, resource optimization in power system engineering, control and optimization of microgrids and smart grids, and optimization issues in power systems and smart grids. He is a reviewer of renowned international journals, such as IEEE Access, IEEE Transactions on Biomedical Engineering, Wiley International Transaction on Electrical Energy Systems, and Journal of Network and Computer Applications.
\end{IEEEbiography}


\begin{IEEEbiography}
[{\includegraphics[width=1in,height=1.25in,clip,keepaspectratio]{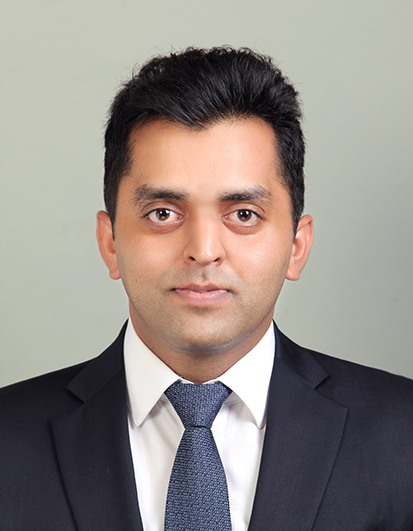}}]
   {Waqas Khalid} (Member, IEEE) received the B.S. degree in Electronics Engineering from GIK Institute of Engineering Sciences and Technology, KPK, Pakistan, in 2011. He received M.S. and Ph.D. degrees in information and communication engineering from Inha University, Incheon, South Korea, and Yeungnam University, Gyeongsan, South Korea, in 2016, and 2019, respectively. He is currently working as a research professor at the Institute of Industrial Technology, Korea University, Sejong, South Korea. His areas of interest include physical layer modeling, signal processing, and emerging technologies for 5G networks, including reconfigurable intelligent surfaces, physical-layer security, NOMA, cognitive radio, UAVs, and IoTs. 
\end{IEEEbiography}

\begin{IEEEbiography}
[{\includegraphics[width=1in,height=1.25in,clip,keepaspectratio]{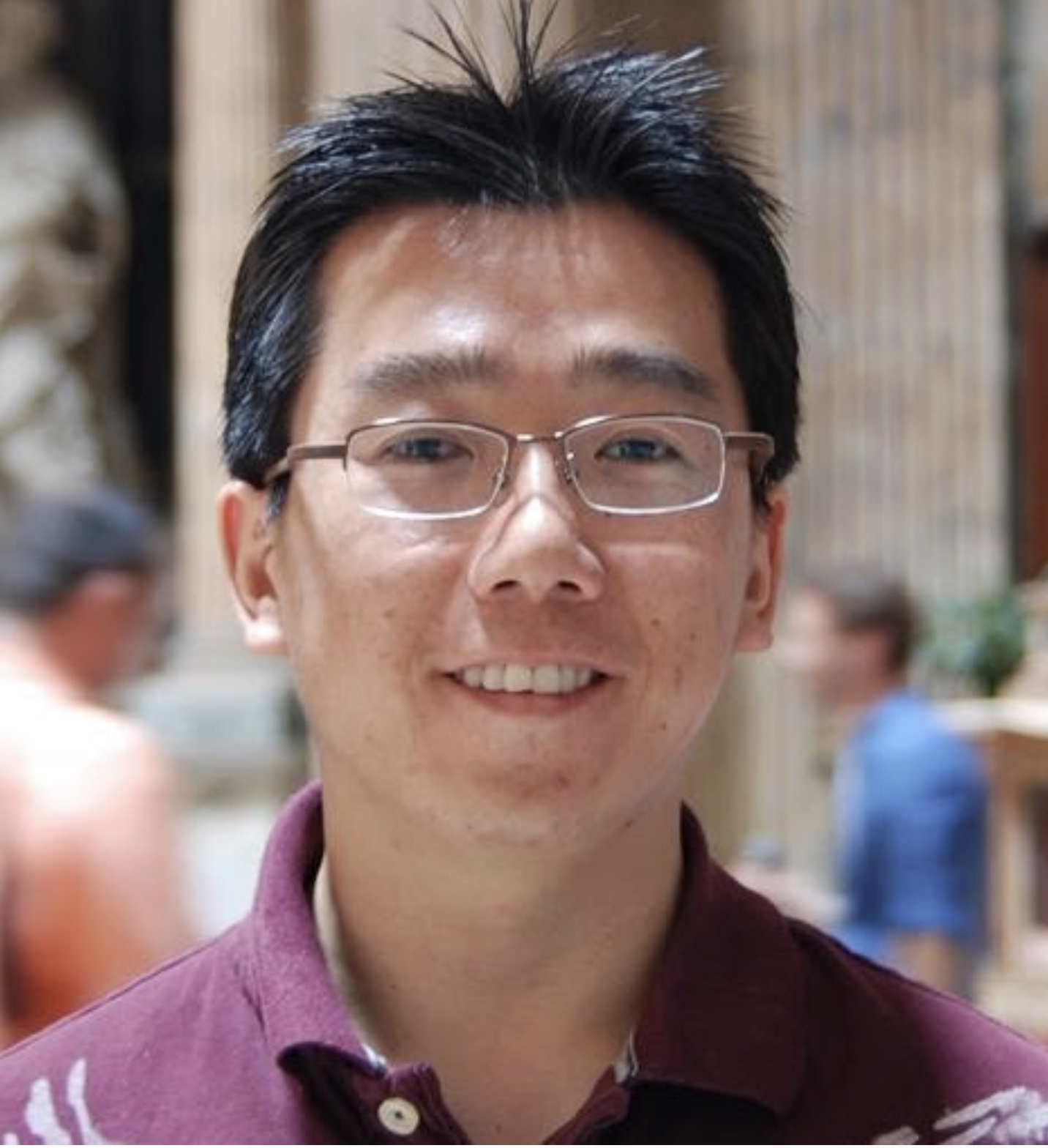}}] {Heejung Yu} (Senior Member, IEEE) received the B.S. degree in radio science and engineering from Korea University, Seoul, South Korea, in 1999, and the M.S. and Ph.D. degrees in electrical engi- neering from the Korea Advanced Institute of Science and Technology, Daejeon, South Korea, in 2001 and 2011, respectively. From 2001 to 2012, he was with the Electronics and Telecommunications Research Institute, Daejeon, and from 2012 to 2019, he was with Yeungman University, Gyeongsan, South Korea. He is currently a Professor with the Department of Electronics and Information Engineering, Korea University, Sejong, South Korea. His research interests include statistical signal processing and communication theory.
\end{IEEEbiography}

\EOD

\end{document}